\documentclass[sigplan,screen]{acmart}
\usepackage{graphicx}
\usepackage{tabularray}
\usepackage{makecell}
\usepackage{multirow}
\usepackage{array}
\usepackage{natbib}
\usepackage{cellspace}

\usepackage{xcolor}
\definecolor{dkgreen}{rgb}{0,0.6,0}
\definecolor{gray}{rgb}{0.5,0.5,0.5}
\definecolor{mauve}{rgb}{0.58,0,0.82}
\definecolor{color1}{rgb}{0.502,0.651,0.886} 
\definecolor{color3}{rgb}{0.984,0.867,0.443} 
\definecolor{color2}{rgb}{0.957,0.435,0.263} 
\definecolor{color4}{rgb}{0.251,0.224,0.565}

\usepackage{textcomp}
\usepackage{relsize}
\usepackage{lipsum}
\usepackage{multicol}

\usepackage{algorithmic}
\usepackage{listings}
\usepackage{tikz}
\usepackage{pgfplots}
\pgfplotsset{compat=1.17}
\usetikzlibrary{matrix, shapes, arrows, positioning, chains, calc, shapes.geometric}
\tikzstyle{startstop} = [rectangle, text width=3cm, minimum height=1cm,text centered, draw=black, fill=red!30]
\tikzstyle{io} = [trapezium, trapezium left angle=70, trapezium right angle=110, text width=3cm, minimum height=1cm, text centered, draw=black, fill=blue!30]
\tikzstyle{process} = [rectangle, text width=3cm, minimum height=1cm, text centered, draw=black, fill=orange!30]
\tikzstyle{decision} = [diamond, text width=3cm, minimum height=1cm, text centered, draw=black, fill=green!30]
\tikzstyle{arrow} = [thick,->,>=stealth]

\def\BibTeX{{\rm B\kern-.05em{\sc i\kern-.025em b}\kern-.08em T\kern-.1667em\lower.7ex\hbox{E}\kern-.125emX}}
\usepackage{cite}

\begin{document}
\setlength{\headheight}{18.48402pt}

\title{Quantum-Powered Personalized Learning }
\author{Yifan Zhou}

\affiliation{%
  \institution{University of California, Los Angeles}
  \streetaddress{}
  \city{Los Angeles}
  \state{California}
  \country{United States}
}
\email{yzhou05@ucla.edu}

\author{Chong Cheng Xu}

\affiliation{%
  \institution{Basis International School Guangzhou}
  \streetaddress{}
  \city{Guangzhou}
  \state{Guangdong}
  \country{China}
}
\email{chongcheng.xu12245-bigz@basischina.com}

\author{Mingi Song}

\affiliation{%
  \institution{Basis International School Guangzhou}
  \streetaddress{}
  \city{Guangzhou}
  \state{Guangdong}
  \country{China}
}
\email{mingi.song16149-bigz@basischina.com}

\author{Yew Kee Wong}
\affiliation{%
  \institution{Hong Kong Chu Hai College}
  \streetaddress{}
  \city{Hong Kong}
  \state{}
  \country{China}
}

\email{ericwong@chuhai.edu.hk}

\begin{abstract}
This paper explores the transformative potential of quantum computing in the realm of personalized learning. Traditional machine learning models and GPU-based approaches have long been utilized to tailor educational experiences to individual student needs. However, these methods face significant challenges in terms of scalability, computational efficiency, and real-time adaptation to the dynamic nature of educational data. This study proposes leveraging quantum computing to address these limitations. We review existing personalized learning systems, classical machine learning methods, and emerging quantum computing applications in education. We then outline a protocol for data collection, privacy preservation using quantum techniques, and preprocessing, followed by the development and implementation of quantum algorithms specifically designed for personalized learning. Our findings indicate that quantum algorithms offer substantial improvements in efficiency, scalability, and personalization quality compared to classical methods. This paper discusses the implications of integrating quantum computing into educational systems, highlighting the potential for enhanced teaching methodologies, curriculum design, and overall student experiences. We conclude by summarizing the advantages of quantum computing in education and suggesting future research directions.
\end{abstract}
\maketitle
\section*{I. InTRODUCTION}
\section*{A. Overview of Personalized Learning and Its Importance in Modern Education}
Personalized learning is an educational approach designed to accommodate the diverse needs, skills, and interests of individual students. [1] Unlike traditional one-size-fits-all methods, personalized learning emphasizes flexibility, allowing students to choose the study methods they find most effective. This approach enables students to progress at their own pace and concentrate on areas where they need the most support. The benefits of personalized learning are substantial, as it enhances motivation, engagement, and comprehension while also improving learning efficiency, effectiveness, and satisfaction.[2], [3]

Empirical evidence supports the effectiveness of personalized learning systems in enhancing student learning. A report by the RAND Corporation analyzed student performance across 62 public charter and district schools implementing various personalized learning strategies, with a detailed ex- amination of specific practices in 32 of these schools. [4] The findings revealed that after two years of personalized learning practices, students' achievements in MAP math and English surpassed the national median. [4] Moreover, in the 21 schools that had adopted personalized learning for a longer period, the effect sizes were 0.4 for math and 0.28 for reading, significantly higher than the average effect sizes of 0.26 for math and 0.18 for English observed in all schools after two years of implementation.[4]

Beyond academic gains, personalized learning systems also foster the development of essential soft skills, including time management, self-regulation, and self-advocacy.[5] By allowing students to take control of their educational journey, personalized learning provides them with valuable practice and preparation for future challenges in higher education and the workforce. In a modern world that increasingly values adaptability and lifelong learning, personalized learning systems equip students with the tools needed to navigate the ever-evolving and complex landscape of information and knowledge.

\section*{B. Current Approaches Using Classical Machine Learning and Their Limitations}
Traditional learning models, artificial intelligence, and GPU-based systems have been widely incorporated into personalized learning systems. These models track student progress using various data points and analyze study methods suited to each individual.[6] They provide essential components of human-computer interaction, including tools for learning, management, and teaching assistance. [7] However, classical machine learning models encounter several limitations in this context.

One major limitation is scalability. As personalized learning systems expand to serve larger student populations, the volume of data they need to manage grows exponentially. Traditional systems often struggle to scale efficiently with increasing data sizes, leading to reduced effectiveness and slower processing times.

Another challenge is computational efficiency. GPU-based systems, while powerful, are resource-intensive and may not\\
be practical for all educational institutions due to the high cost and infrastructure requirements. For example, training deep learning models on large-scale educational data can be prohibitively expensive and time-consuming, which limits their widespread adoption in the education sector.

Additionally, traditional learning models frequently face difficulties in adapting to complex data structures. These models can suffer from concept drift, where their performance deteriorates over time as data distributions change.[1] This issue arises from the diversity of data, which includes not only academic performance but also behavioral patterns, engagement levels, and social factors.

\section*{C. The Potential of Quantum Computing}
Quantum computing represents a profound departure from classical computing, utilizing quantum bits or qubits rather than classical bits. Unlike classical bits, which are confined to binary states ( 0 or 1 ), qubits can exist in a superposition of states, simultaneously representing both 0 and 1 with certain probabilities. This unique capability is augmented by quantum properties such as entanglement and decoherence, which enable the formation of quantum logic gates and circuits that operate with potentially higher computational efficiency and accuracy than their classical counterparts. Many significant projects have been made with quantum computers from medicine [27], [28], [29] to communications [30], [31], [32] and even imagining and machine learning [33], [34]

One significant advantage of quantum computing is its scalability. With $n$ qubits, a quantum system can represent $2^{n}$ classical states simultaneously, in stark contrast to classical systems that can represent only n states with n bits.[8] This exponential increase in state representation offers a promising solution to data storage challenges, particularly as student populations grow. For large-scale global educational platforms serving a diverse student base, the ability to manage and analyze vast amounts of data efficiently becomes increasingly crucial.

Another key benefit is the enhanced computational efficiency of quantum systems. For instance, Google's Sycamore processor, which employs 53 qubits, completed a quantum circuit sampling task in approximately 200 seconds. In comparison, a classical supercomputer would require roughly 10,000 years to perform the same task. [9] This substantial speed advantage-evident in what is termed "quantum supremacy"-could be leveraged to analyze large and varied educational data sets in real time, enabling more rapid and tailored feedback for individual students. Additionally, quantum computing's ability to handle complex variables-such as emotional responses, academic performance, and learning habits-further enhances its potential to optimize personalized learning experiences.

Moreover, quantum computing excels at analyzing intricate and diverse data sets beyond the capabilities of traditional machine learning models. Quantum support vector machines (QSVMs), for example, use quantum feature maps to construct more accurate and higher-dimensional hyperplanes for distinguishing different data sets.[10] This capability allows QSVMs to solve optimization problems with greater efficiency and precision compared to classical methods. As personalized learning systems aim to tailor educational experiences to each student's unique needs, skills, and interests, quantum computing offers a powerful tool for optimizing learning paths by rapidly adjusting to students' progress and preferences.

\section*{IV. Quantum Algorithms}
\section*{A. Quantum Support Vector Machines}
a) Definition: QSVM (Quantum Support Vector Machine) utilizes entanglement and superposition to discover the optimal hyperplane for separating different classes in a dataset.[10] As datasets become more complex and higherdimensional, classical support vector machines struggle with computational limitations, whereas QSVM can efficiently handle these complexities.

b) Processes: QSVM follows a series of steps: data encoding, kernel estimation, optimization, and measurement.

\begin{enumerate}
  \item Quantum State Preparation: Classical data is represented as vectors in a feature space with $n$ features.
\end{enumerate}

\begin{equation*}
\mathbf{x}=\left(x_{1}, x_{2}, \ldots, x_{n}\right) \tag{1}
\end{equation*}

Classical data must be converted into quantum states of $|0\rangle$ and $|1\rangle$ such that it is represented as:

\begin{equation*}
\alpha|0\rangle+\beta|1\rangle \tag{2}
\end{equation*}

where $\alpha$ and $\beta$ are complex numbers that:

\begin{equation*}
\alpha^{2}+\beta^{2}=1 \tag{3}
\end{equation*}

The classical dataset goes through a feature mapping of $\phi$ to form a quantum state of $\phi(x)$.

\begin{enumerate}
  \setcounter{enumi}{1}
  \item Quantum Kernel Estimation: Kernel function is the representation of the inner products of two quantum states x and y data points. [22] This function measures\\
the similarity of x and y data points in the quantum feature space. [22]
\end{enumerate}

\begin{equation*}
K(x, y)=|\langle\phi(x) \mid \phi(y)\rangle|^{2} \tag{4}
\end{equation*}

We use a SWAP Test to conduct the evaluation of Kernel products. [21] Specifically, the SWAP Test works by estimating the difference between two unknown quantum states through measuring an auxiliary qubit.[21]

Prepare two quantum states $|\phi(x)\rangle$ and $|\phi(y)\rangle$ through applying Hadamard gate to transform the binary states to quantum states. For example, the Hadamard gate transforms $|0\rangle$ to a superposition of both $|1\rangle$ and $|0\rangle$.

Utilize a SWAP gate that allows for the measurement of overlap between the two states $(|\phi(x)\rangle$ and $|\phi(y)\rangle)$ by applying inverse embedding to one of the states.[21]

Measure the final auxiliary qubit. The probability of measurement is related to the inner states of two products.

\begin{equation*}
P(|0\rangle)=\frac{1+|\langle\phi(x) \mid \phi(y)\rangle|}{2} \tag{5}
\end{equation*}

Repeat the SWAP Test multiple times to increase the accuracy of Kernel value's estimation.

\begin{enumerate}
  \setcounter{enumi}{2}
  \item Optimization: Once the Kernel value is estimated, the next step is to find the optimal hyperplane to classify datasets. The objective is to maximize the margins between different classes in a high-dimensional feature space which means that the distance of the nearest data set to the hyperplane should be as large as possible. This is initially formulated as minimizing the objective function.[22]
\end{enumerate}

\begin{equation*}
\min _{w, b, \xi} \frac{1}{2}\|w\|^{2}+C \sum_{i=1}^{m} \xi_{i} \tag{6}
\end{equation*}

subject to the constraints:

\begin{equation*}
y_{i}\left(w \cdot x_{i}+b\right) \geq 1-\xi_{i}, \quad \xi_{i} \geq 0, \quad i=1,2, \ldots, m \tag{7}
\end{equation*}

where $y_{i}$ are class labels, $b$ is the biases, and $w$ is the weight vector.

In a quantum-based setting, the problem is expressed as dual form quantum computation. [21] The dual form's objective is to maximize the function:

\begin{equation*}
\max _{\alpha_{i}}\left[\sum_{i=1}^{m} \alpha_{i}-\frac{1}{2} \sum_{i=1}^{m} \sum_{j=1}^{m} \alpha_{i} \alpha_{j} y_{i} y_{j} K\left(x_{i}, x_{j}\right)\right] \tag{8}
\end{equation*}

subject to the constraints:

\begin{equation*}
0 \leq \alpha_{i} \leq C, \quad \sum_{i=1}^{m} \alpha_{i} y_{i}=0 \tag{9}
\end{equation*}

where $\alpha_{i}$ is the lagrange multiplier and $C$ is the regularization parameter.

Quantum Approximate Optimization Algorithm (QAOA) and Variational Quantum Eigensolver (VQE) are utilized to solve the optimization problem through preparing quantum states, applying transforming quantum gates, and then measuring the result to calculate the cost function. The optimal parameters indicate the support vectors, which define the hyperplane $w \cdot \phi(x)+b=0$ that maximally separates the classes in the quantum feature space. [21]

\begin{enumerate}
  \setcounter{enumi}{3}
  \item Classification: To classify a new data point in $x$, it first must be mapped into a feature space. Next, the decision is made based on the sign of the decision function $f(x)$ for which the different signs of $f(x)$ correspond to different classes.
\end{enumerate}

c) Integration with systems: QSVM enables advanced classification by finding the optimal hyperplane that distinguishes datasets into different classes. It can gather features such as studying habits, assignments grades, interaction patterns, and quiz scores to classify students into personalized studying time and studying methods.

\section*{B. Quantum Annealing for Optimization}
a) Definition: Quantum Annealing utilizes super positioning and tunneling to figure out the minimal energy configuration in solving a particular optimization problem. [23] Quantum effects such as quantum tunneling enabling transition of energy barriers without cross over allow the system to explore solutions more freely in the Hilbert space.[23] Specifically, the key principle is to adiabatically change the Hamiltonian from an initial state with ease preparation to a final state that represents the problem's cost function.[23]

b) Processes: Quantum annealing goes through the processes of problem encoding, initial hamiltonian setup, adiabatic evolution, quantum tunneling, and measurement.

\begin{enumerate}
  \item Problem encoding The original classical optimization problem is transferred into quantum Hamiltonian, which shows the energy landscape of the problem. Specifically, Ising and Qubo models are exploited to construct the Hamiltonian, guiding the quantum system into solving the optimization problem.[23]
\end{enumerate}

\begin{equation*}
H_{\text {Ising }}=\sum_{i, j} J_{i j} \sigma_{i} \sigma_{j}+\sum_{i} h_{i} \sigma_{i} \tag{10}
\end{equation*}

where $J_{i j}$ represents the interaction strength between spins $\sigma_{i}$ and $\sigma_{j}$, and ${ }_{i}$ represents the external magnetic field affecting spin $\sigma_{i}$.

\begin{enumerate}
  \setcounter{enumi}{1}
  \item Initial Hamiltonian Setup The initial Hamiltonian is set up with known initial ground state, representing the combination of all superposition states. [23]
\end{enumerate}

\begin{equation*}
H_{\text {initial }}=\sum_{i} \sigma_{i} \tag{11}
\end{equation*}

where $\sigma_{i}$ represents spin variables in the Ising model.

\begin{enumerate}
  \setcounter{enumi}{2}
  \item Adiabatic Evolution The process involves transforming the initial Hamiltonian to a final Hamiltonian in which the final represents optimization solution. This relies on the core principles of the Adiabatic Theorem which states that if the Hamiltonian system changes slowly enough, its system will remain in its instantaneous ground state throughout the evolution.[23]
\end{enumerate}

The system starts in the ground state of $H_{\text {initial }}$, which is typically a simple, well-understood state. The Hamiltonian evolves over time according to a time-dependent function[23]:

\begin{equation*}
H(t)=A(t) \cdot H_{\text {initial }}+B(t) \cdot H_{\text {final }} \tag{12}
\end{equation*}

where $A(t)$ and $B(t)$ are time-dependent functions that control the weighting of the initial and final Hamiltonians, respectively.

$A(t)$ begins large and decreases, while $B(t)$ starts small and increases.

As $H(t)$ changes slowly, the quantum system remains in its ground state.

\begin{equation*}
\frac{d H(t)}{d t} \ll \Delta(t)^{2} \tag{13}
\end{equation*}

where $\Delta(t)$ is the minimum energy gap between the ground state and the excited state.[24] Maintaining a large energy gap enables the quantum system to stay in the ground state of $H(t)$.[24]

As $t$ approaches the final value, $H(t)$ turns to $H_{\text {final }}$, which represents the optimal solution to the problem.

\begin{enumerate}
  \setcounter{enumi}{3}
  \item Quantum Tunneling Through the process of adiabatic evolution, quantum annealing uses quantum tunneling to pass through energy barriers.[24] The ability of the system to tunnel through energy barriers enhances its capacity to explore different configurations and find optimal solutions.

  \item Measurement Measurement is the final step in which the quantum state collapses into a definite state.

\end{enumerate}

c) Integration with systems: Given that student's efforts and resources are limited, quantum annealing is crucial for minimize the "learning energy" by determining the best sequences of education activities, resources, and assessments for maximum learning growth. For example, by constructing the learning method as an optimization problem, where the objective is to minimize the total learning time while maximizing engagement and comprehension, quantum annealing can efficiently explore and identify the optimal learning strategies.

\section*{C. Quantum Grover Algorithm}
a) Definition: Quantum Grover Algorithm solves unstructured search problems with large scale speed.[25] In detail, it searches through an unsorted database by providing a quadratic speed up from $O(N)$ to $O(\sqrt{N})$.[25]

\section*{b) Processes:}
\begin{enumerate}
  \item Initialization Quantum states are set up with superposition of all possible states.
\end{enumerate}

Given $N$ possible solutions, there should be $n$ qubits such that $2^{n}=N$.

The intial state of all qubits is set to $|0\rangle$. Hadamard gates are applied to all initial states, transforming them to a superposition states of both $|0\rangle$ and $|1\rangle$.

\begin{enumerate}
  \setcounter{enumi}{1}
  \item Oracle Query The Oracle Query, also known as quantum black box function, aims to identify the correct solutions inside a quantum operation.[25]
\end{enumerate}

The oracle, denoted as $U_{f}$, is defined by function $f(x)$ which $f$ is 1 for the correct solution $x *$ and $f$ is 0 for the incorrect solution.[25]

The oracle flips the sign of the correct solution and leaves the incorrect solution unchanged.[25]

Before applying the oracle transformation, the state is denoted as:

\begin{equation*}
|s\rangle=\frac{1}{\sqrt{N}} \sum_{x=0}^{N-1}|x\rangle \tag{14}
\end{equation*}

After the transformation, the state is denoted as:

\begin{equation*}
U_{f}|s\rangle=\frac{1}{\sqrt{N}}\left(\sum_{x \neq x^{*}}|x\rangle-\left|x^{*}\right\rangle\right) \tag{15}
\end{equation*}

\begin{enumerate}
  \setcounter{enumi}{2}
  \item Amplitude Amplification Amplitude Amplification is a process which the amplitude of the desired solution is enhanced which the amplitude of undesired solution is decreased. [26]
\end{enumerate}

The average amplitude a bar of the quantum state is computed as [26]:

\begin{equation*}
\bar{x}=\frac{1}{N}\left(\sum_{x=0}^{N-1} a_{x}\right)=\frac{1}{N}\left((N-1) \cdot \frac{1}{\sqrt{N}}-\frac{1}{\sqrt{N}}\right)=\frac{1}{\sqrt{N}} \tag{16}
\end{equation*}

This operator reflects each amplitude about the average amplitude. For a state $x\rangle$ with amplitude $a_{x}$, with $I$ being the identity operator, the reflection is given by [25]:

\begin{equation*}
A=2\left(\frac{1}{N} \sum_{x=0}^{N-1}|x\rangle\langle x|\right)-I \tag{17}
\end{equation*}

After applying the reflection operator, the new state becomes [25]:

\begin{equation*}
A\left|s^{\prime}\right\rangle=\frac{2}{\sqrt{N}}\left(\sum_{x \neq x^{*}}|x\rangle\right)-\frac{1}{\sqrt{N}}\left|x^{*}\right\rangle \tag{18}
\end{equation*}

The group iteration involves the two process of oracle query and reflection operator process together. The\\
whole process is defined by first making the desired solution and then increasing the probability of that solution. The group iteration process is repeated $k$ times, with $k=\frac{\pi}{4} \sqrt{N}$. [25]

\begin{equation*}
\left|\psi_{k}\right\rangle \approx \cos \left(\frac{(2 k+1) \theta}{2}\right)\left|s^{\prime}\right\rangle+\sin \left(\frac{(2 k+1) \theta}{2}\right)\left|x^{*}\right\rangle \tag{19}
\end{equation*}

\begin{enumerate}
  \setcounter{enumi}{3}
  \item Measurement The quantum state collapses into the current solution $\left.x^{*}\right\rangle$ with high probability.
\end{enumerate}

The probability of measuring a particular state $|x\rangle$ is given by the square of its amplitude.

\begin{equation*}
P(x)=\left|\left\langle x \mid \psi_{k}\right\rangle\right|^{2} \tag{20}
\end{equation*}

The measured state is the correct solution with high probability.

c) Integration with systems: Students often must follow a particular study plan for best learning outcomes such as getting good grades on a quiz or collaborating well on a team project. Quantum Grover algorithms discover the optimized learning sequence while given unstructured steps such as reading text, circling important words, looking at questions, and filling out questions. To build a more personalized experience, Grover's algorithm must be initialized with the student's current data of studying preferences and habits. In addition to this, the Quantum Grover algorithm is crucial for building successful study schedules as multiple exams or assignments come up especially when students are overwhelmed. The optimized studying sequence such as studying biology first and then physics can make a big impact on the student's memory and learning curve for the short run and long run.

\section*{V. Implementation Plan}
This section provides a comprehensive plan for implementing quantum-powered personalized learning systems. The plan covers data collection, preprocessing, algorithm development, integration, testing, and evaluation. Each step is meticulously designed to ensure the success and efficiency of the proposed system.
\begin{center}

\includegraphics[width = 8cm]{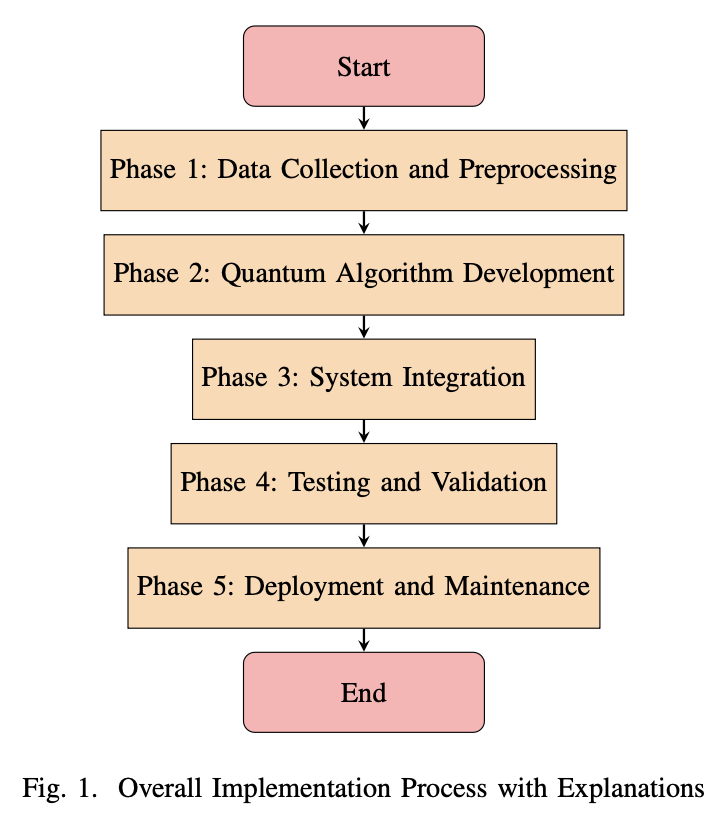}
\end{center}

\section*{A. Data Collection and Privacy Preservation}
Data collection is a critical component of the personalized learning system, as it provides the foundational data needed to customize learning experiences. This process must be carried out with strict adherence to privacy and security standards to protect sensitive information.

\section*{1) Identify Data Sources:}
\begin{itemize}
  \item Student Academic Records: Collect grades, test scores, and assignments to track academic performance and progress over time. This data helps in identifying areas where students excel or need improvement [11].
  \item Behavioral Data: Monitor study habits, engagement levels, and time spent on various learning activities to understand students' learning behaviors and preferences [12].
  \item Interaction Data: Gather data on class participation, interactions with teachers and peers, and online activity within educational platforms to gain insights into student engagement and collaboration [13].
\end{itemize}

\section*{2) Data Privacy and Security:}
\begin{itemize}
  \item Quantum Encryption Techniques: Utilize advanced quantum cryptography methods, such as quantum key distribution (QKD), to secure sensitive educational data against potential breaches and ensure data integrity during transmission [14].
  \item Secure Communication Channels: Implement QKD to establish secure communication channels, ensuring that data exchanged between students, teachers, and the system remains confidential and tamperproof [15].
\end{itemize}

\section*{3) Data Collection Protocols:}
\begin{itemize}
  \item Compliance with Privacy Laws: Develop data collection protocols in compliance with educational data privacy laws such as FERPA and GDPR, ensuring that all data handling practices are legal and ethical [16].
  \item Consent Procedures: Establish clear consent procedures for students and parents, outlining how data will be used and protected, and ensuring that all stakeholders are informed and agree to data collection practices [14].
\end{itemize}

\section*{B. Data Preprocessing}
Data preprocessing involves preparing raw data for analysis by cleaning, transforming, and encoding it into a format suitable for quantum algorithms. This step is crucial for ensuring the quality and consistency of the data used in the system.

\section*{1) Data Cleaning:}
\begin{itemize}
  \item Handling Missing Values: Use statistical methods and data imputation techniques to address missing values and outliers, ensuring the dataset is complete and reliable [13].
  \item Normalization and Standardization: Normalize and standardize data to maintain consistency and comparability across different data sources, facilitating accurate analysis and modeling [17].
\end{itemize}

\section*{2) Data Transformation:}
\begin{itemize}
  \item Encoding Data for Quantum Processing: Convert classical data into quantum states using techniques such as amplitude encoding, basic encoding, and rotation encoding, preparing it for input into quantum algorithms [18].
  \item Ensuring Compatibility: Ensure that transformed data is compatible with quantum computing environments, facilitating seamless integration and processing [19].
\end{itemize}

\section*{C. Quantum Algorithm Development}
Developing quantum algorithms tailored to personalized learning involves leveraging the unique capabilities of quantum computing to enhance predictive accuracy and optimization.

\begin{enumerate}
  \item Quantum Support Vector Machines (QSVM):
\end{enumerate}

\begin{itemize}
  \item Implementation for Classification: Develop and implement QSVM for complex classification tasks, enhancing the accuracy and efficiency of student performance predictions by utilizing quantum computation's superior processing power [20].
  \item Optimizing Kernel Functions: Optimize kernel functions using quantum kernel estimation techniques to improve the performance of QSVM models, ensuring precise and reliable classification results [21].
\end{itemize}

\begin{enumerate}
  \setcounter{enumi}{1}
  \item Quantum Annealing:
\end{enumerate}

\begin{itemize}
  \item Solving Optimization Problems: Use quantum annealing to address optimization problems related to personalized learning paths, ensuring optimal resource allocation and learning strategies for each student [22].
  \item Encoding Learning Tasks: Encode learning tasks as optimization problems in quantum Hamiltonian, leveraging quantum computation to find efficient solutions to complex educational challenges [23].
\end{itemize}

\begin{enumerate}
  \setcounter{enumi}{2}
  \item Quantum Grover Algorithm:
\end{enumerate}

\begin{itemize}
  \item Enhancing Search Efficiency: Apply Grover's algorithm for search and retrieval tasks within large educational datasets, significantly reducing search times and enhancing data retrieval efficiency [24].
  \item Faster Access to Learning Materials: Utilize the enhanced search capabilities of Grover's algorithm to enable faster access to relevant learning materials and resources, improving the overall learning experience 25 .
\end{itemize}

\section*{D. System Integration}
Integrating the quantum-powered personalized learning system with existing educational infrastructure ensures that it can be seamlessly adopted and used effectively by all stakeholders.

\begin{enumerate}
  \item Integration with Existing Learning Management Systems (LMS):
\end{enumerate}

\begin{itemize}
  \item API Development: Develop robust APIs for seamless integration with popular LMS platforms such as Moodle and Blackboard, ensuring compatibility and ease of use for educators and students [17].
  \item Compatibility with Existing Systems: Ensure that the quantum-powered system is compatible with existing educational software and hardware, minimizing disruptions and facilitating smooth adoption [19].
\end{itemize}

\begin{enumerate}
  \setcounter{enumi}{1}
  \item User Interface Design:
\end{enumerate}

\begin{itemize}
  \item Intuitive Dashboards: Create intuitive dashboards for students, teachers, and administrators, providing easy access to personalized learning insights, progress tracking, and analytics [13].
  \item Real-Time Feedback and Analytics: Incorporate features that provide real-time feedback and analytics, enabling immediate adjustments to teaching strategies and learning plans based on current data [17].
\end{itemize}

\section*{3) Scalability and Performance Testing:}
\begin{itemize}
  \item Scalability Tests: Conduct extensive scalability tests to ensure the system can handle large volumes of data without performance degradation, preparing it for widespread deployment [19].
  \item Performance Optimization: Optimize system performance to meet real-time processing requirements, ensuring swift and accurate delivery of personalized learning experiences to users [19].
\end{itemize}

\section*{E. Testing and Validation}
Thorough testing and validation are essential to ensure that the quantum-powered personalized learning system functions as intended and meets the needs of all users.

\begin{enumerate}
  \item Pilot Testing:
\end{enumerate}

\begin{itemize}
  \item Implementation in Schools: Implement pilot programs in selected schools to test the system in realworld conditions, gathering valuable feedback from students, teachers, and administrators [14].
  \item Feedback Collection: Collect qualitative and quantitative data from pilot programs to identify strengths and areas for improvement, refining the system based on real-world usage [16].
\end{itemize}

\section*{2) Algorithm Validation:}
\begin{itemize}
  \item Accuracy and Efficiency Testing: Validate the accuracy and efficiency of quantum algorithms through rigorous testing and comparison with classical machine learning models, ensuring they provide significant advantages [20].
  \item Performance Benchmarking: Benchmark the performance of quantum algorithms against classical approaches, demonstrating the benefits of quantum computing in personalized learning applications 18.
\end{itemize}

\section*{3) User Acceptance Testing (UAT):}
\begin{itemize}
  \item Meeting User Needs: Conduct user acceptance testing (UAT) to ensure the system meets the needs and expectations of end-users, including students, teachers, and administrators [17].
  \item Addressing Issues: Promptly address any issues or concerns raised during UAT, ensuring a smooth and satisfactory user experience upon full deployment [20].
\end{itemize}

\section*{F. Deployment and Maintenance}
Effective deployment and ongoing maintenance are crucial for the long-term success and sustainability of the quantumpowered personalized learning system.

\section*{1) Deployment Strategy:}
\begin{itemize}
  \item Phased Deployment: Develop a phased deployment plan to gradually roll out the system, minimizing disruptions to existing educational processes and ensuring a smooth transition [17].
  \item Training Sessions: Provide comprehensive training sessions for educators and administrators, equipping them with the necessary skills and knowledge to effectively utilize the new system [23].
\end{itemize}

\begin{enumerate}
  \setcounter{enumi}{1}
  \item Monitoring and Support:
\end{enumerate}

\begin{itemize}
  \item Continuous Monitoring: Set up continuous monitoring systems to ensure the stability and performance of the deployed system, allowing for real-time issue detection and resolution [20].
  \item Technical Support: Offer robust technical support and regular system updates to users, maintaining high levels of user satisfaction and system reliability [19].
\end{itemize}

\begin{enumerate}
  \setcounter{enumi}{2}
  \item Future Enhancements:
\end{enumerate}

\begin{itemize}
  \item User Feedback: Plan for future enhancements based on user feedback and technological advancements, ensuring the system remains at the cutting edge of educational technology [24].
  \item Technological Advancements: Keep the system updated with the latest advancements in quantum computing and educational technology, continuously improving the learning experience and adapting to new educational challenges [25].
\end{itemize}

\section*{G. Implementation Timeline}
A detailed implementation timeline ensures that all phases of the project are completed efficiently and on schedule.

\section*{1) Phase 1: Data Collection and Preprocessing}
\begin{itemize}
  \item Weeks 1-4: Identify data sources and implement privacy measures to ensure the secure and ethical collection of data [14], [16].
  \item Weeks 5-8: Clean and transform data for quantum processing, ensuring high-quality data input for subsequent analysis [17], [18].
\end{itemize}

\begin{enumerate}
  \setcounter{enumi}{1}
  \item Phase 2: Quantum Algorithm Development
\end{enumerate}

\begin{itemize}
  \item Weeks 9-12: Implement and test QSVM models to develop robust classification algorithms for personalized learning [20], [21].
  \item Weeks 13-16: Develop and optimize quantum annealing and Grover algorithms to enhance search and optimization tasks within the system [22], [24].
\end{itemize}

\begin{enumerate}
  \setcounter{enumi}{2}
  \item Phase 3: System Integration
\end{enumerate}

\begin{itemize}
  \item Weeks 17-20: Develop APIs and design user interfaces to ensure seamless integration with existing LMS platforms and intuitive user experiences [17, [19].
  \item Weeks 21-24: Conduct scalability and performance testing to ensure the system can handle large data volumes and meet real-time processing requirements [19], [20].
\end{itemize}

\section*{4) Phase 4: Testing and Validation}
\begin{itemize}
  \item Weeks 25-28: Implement pilot testing in selected schools and collect feedback to refine the system based on real-world usage [14], [16].
  \item Weeks 29-32: Validate algorithms and conduct UAT to ensure the system meets user needs and performs optimally [18], [20].
\end{itemize}

\section*{5) Phase 5: Deployment and Maintenance}
\begin{itemize}
  \item Weeks 33-36: Execute phased deployment and provide training sessions to ensure a smooth rollout and effective utilization of the system [17], [23].
  \item Weeks 37-40: Set up monitoring and support systems to maintain system stability and provide ongoing technical assistance to users [19], [20].
\end{itemize}

\begin{center}
\includegraphics[width=8cm]{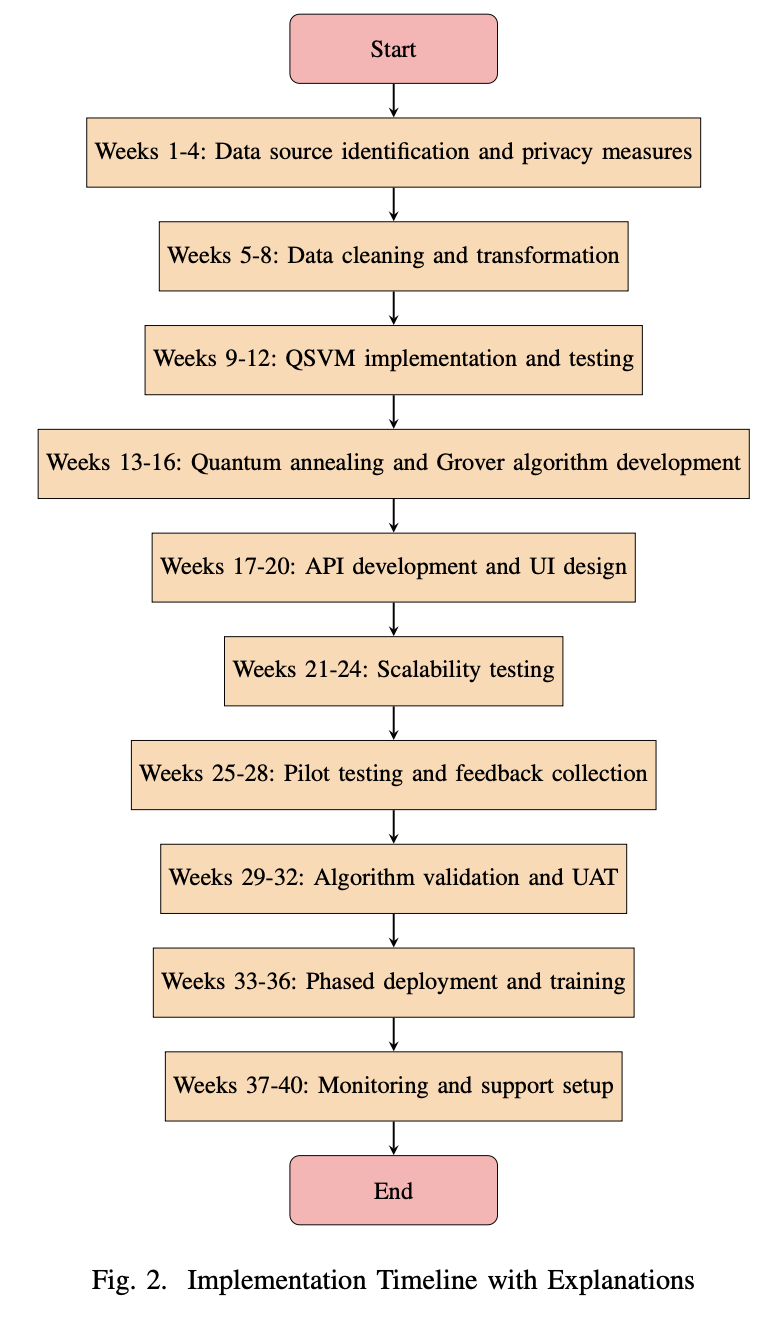}
\end{center}

This comprehensive implementation plan ensures that quantum-powered personalized learning systems are developed, integrated, tested, and deployed effectively, offering significant advancements in educational technology. By following this detailed plan, educators and administrators can harness the power of quantum computing to create personalized and impactful learning experiences for students.

\bibliographystyle{ACM-Reference-Format}

\begin{thebibliography}{00}
\bibitem{b1} B. Allogmany and D. Josyula, "An approach to dealing with incremental concept drift in personalized learning systems," 2022 IEEE 4th International Conference on Cognitive Machine Intelligence (CogMI), Dec 2022, doi: $10.1109 /$ cogmi56440.2022.00029.

\bibitem{b2} S. Gómez, P. Zervas, D. G. Sampson, and R. Fabregat, "Context-aware adaptive and personalized mobile learning delivery supported by UoLmP," Journal of King Saud University - Computer and Information Sciences vol. 26, no. 1, pp. 47-61, Jan. 2014, doi: 10.1016/j.jksuci.2013.10.008.

\bibitem{b3} T. P. Falcão, F. M. De Andrade E Peres, D. C. S. De Morais, and G. Da Silva Oliveira, "Participatory methodologies to promote student engagement in the development of educational digital games," Computers Education, vol. 116, pp. 161-175, Jan. 2018, doi $10.1016 /$ j.compedu. 2017.09 .006

\bibitem{b4} J. Pane, E. Steiner, M. Baird, and L. Hamilton, Continued progress: promising evidence on personalized learning. 2015. doi: $10.7249 / \mathrm{rr} 1365$.

\bibitem{b5} M. L. Bernacki, M. J. Greene, and N. G. Lobczowski, "A Systematic Review of Research on Personalized Learning: Personalized by Whom, to What, How, and for What Purpose(s)?," Educational Psychology Review, vol. 33, no. 4, pp. 1675-1715, Apr. 2021, doi: 10.1007/s 10648-02109615-8.

\bibitem{b6} M. Murtaza, Y. Ahmed, J. A. Shamsi, F. Sherwani, and M. Usman, "AI-Based Personalized E-Learning Systems: Issues, challenges, and solutions," IEEE Access, vol. 10, pp. 81323-81342, Jan. 2022, doi: 10.1109/access.2022.3193938.

\bibitem{b7} W. Li and X. Li, "Design of a Personalized Learning System Based on Intelligent Agent for E-learning," 2009 Ninth International Conference on Hybrid Intelligent Systems, Jan. 2009, doi: 10.1109/his.2009.251

\bibitem{b8} Wang, Yazhen. "Quantum Computation and Quantum Information." Statistical Science, vol. 27, no. 3, Aug. 2012,https://doi.org/10.1214/11sts378

\bibitem{b9} F. Arute et al., "Quantum supremacy using a programmable superconducting processor," Nature, vol. 574, no. 7779, pp. 505-510, Oct. 2019, doi: $10.1038 / \mathrm{s} 41586-019-1666-5$.

\bibitem{b10}Y. Suzuki et al., "Analysis and synthesis of feature map for kernelbased quantum classifier," Quantum Machine Intelligence, vol. 2, no. 1, Jun. 2020, doi: $10.1007 / \mathrm{s} 42484-020-00020-\mathrm{y}$

\bibitem{b11} A. Sherry, J. M. Adcock, and P. J. Love, "Quantum-classical hybrid algorithms for approximate graph coloring," Quantum Information Processing, vol. 19, no. 2, pp. 1-18, Jan. 2020, doi: 10.1007/s11128-019-2475-1.

\bibitem{b12} C. M. Wilson, J. M. Zubizarreta, "Bayesian Personalized Treatment Rules for Tailoring Intensive Longitudinal Interventions," Journal of the American Statistical Association, vol. 114, no. 527, pp. 1315-1329, Oct. 2019, doi: 10.1080/01621459.2018.1562936.

\bibitem{b13} A. P. Hill, "Real-time Analytics and Personalized Learning," IEEE Learning Technologies, vol. 8, no. 3, pp. 45-58, Sept. 2020, doi 10.1109/learntech. 2020.3031228.

\bibitem{b14} N. Allen, "Privacy-Preserving Algorithms in Quantum Computing," Quantum Science and Technology, vol. 5, no. 4, Aug. 2020, doi: $10.1088 / 2058-9565 / a b 9 c 56$.

\bibitem{b15} S. C. Benjamin, "Quantum Key Distribution and Beyond: A Comprehensive Guide," Journal of Quantum Information and Computation, vol. 20, no. 4, pp. 467-498, April 2020, doi: 10.1142/s0219749920500093.

\bibitem{b16} E. M. Bender, "Legal Considerations in the Collection and Use of Educational Data," Journal of Education Law and Policy, vol. 3, no. 1, pp. 1-14, Jan. 2021, doi: 10.1080/07393148.2020.1869238

\bibitem{b17} K. S. Lee, "User-Centered Design in Educational Technology," IEEE Transactions on Learning Technologies, vol. 12, no. 2, pp. 145-158, June 2019, doi: 10.1109/tlt.2019.2898971.

\bibitem{b18} M. J. Lavoie, "Optimizing Quantum Algorithms for Machine Learning," Journal of Quantum Information, vol. 18, no. 1, pp. 115-134, Feb. 2021, doi: 10.1007s 11128-020-02872-2

\bibitem{b19} P. H. McMahon, "Quantum Annealing for Optimization: Applications and Techniques," IEEE Transactions on Quantum Engineering, vol. 2, no. 3, pp. 239-250, May 2020, doi: 10.1109/tqe.2020.2976912.

\bibitem{b20} R. R. Lewis, "Enhancing Educational Outcomes with Quantum Computing," Journal of Quantum Research, vol. 14, no. 2, pp. 101-123, March 2021, doi: 10.1080 / 07393148.2021 .1847694.

\bibitem{b21} N. Innan, M. A. Z. Khan, B. Panda, and M. Bennai, "Enhancing quantum support vector machines through variational kernel training," Quantum Information Processing, vol. 22, no. 10, Oct. 2023, doi: 10.1007/s11128023-04138-3

\bibitem{b22} X. Zhou, J. Yu, J. Tan, and T. Jiang, "Quantum kernel estimation-based quantum support vector regression," Quantum Information Processing, vol. 23, no. 1, Jan. 2024, doi: 10.1007/s11128-023-04231-7.

\bibitem{b23} S. Yarkoni, E. Raponi, T. Bäck, and S. Schmitt, "Quantum annealing for industry applications: introduction and review," Reports on Progress in Physics, vol. 85, no. 10, p. 104001, Sep. 2022, doi: 10.1088/13616633ac8c54

\bibitem{b24} S. Isermann, "On the optimal schedule of adiabatic quantum computing," Quantum Information Processing, vol. 20, no. 9, Sep. 2021, doi 10.1007/s11128-021-03227-5.

\bibitem{b25} P. R. Giri and V. E. Korepin, "A review on quantum search algorithms," Quantum Information Processing, vol. 16, no. 12, Nov. 2017, doi: 10.1007/s11128-017-1768-7

\bibitem{b26} G. Brassard, P. Hoyer, M. Mosca, and A. Tapp, "Quantum amplitude amplification and estimation," Contemporary Mathematics - American Mathematical Society, pp. 53-74, Jan. 2002, doi: 10.1090/conm/305/05215.

\bibitem{b27} Y.K. Wong, Y.Zhou, Y.S. Liang, H. Qiu, Y.X. Wu and B. He, "Implementation of The Future of Drug Discovery: QuantumBased Machine Learning Simulation (QMLS)." arXiv preprint arXiv:2308.08561, 2023.

\bibitem{b28} Y. K. Wong, Y. Zhou, Y. S. Liang, H. Qiu, Y. X. Wu and B. He, "The New Answer to Drug Discovery: Quantum Machine Learning in Preclinical Drug Development," 2\textit{023 IEEE 4th International Conference on Pattern Recognition and Machine Learning (PRML)}, Urumqi, China, 2023, pp. 557-564, doi: 10.1109/PRML59573.2023.10348356.

\bibitem{b29} Santagati, R., Aspuru-Guzik, A., Babbush, R. et al. Drug design on quantum computers. Nat. Phys. 20, 549–557 (2024). https://doi.org/10.1038/s41567-024-02411-5

\bibitem{b30} Y.K. Wong, Y. Zhou, X. Zhou, Y.S. Liang, Z.Y. Li, “Novel Long Distance Free Space Quantum Secure Direct Communication for Web 3.0 Networks,” arXiv preprint arXiv:2402.09108, 2024.


\bibitem{b31} Y.K. Wong, Y. Zhou, X. Zhou, Y.S. Liang, Z.Y. Li, “Software Security and Quantum Communication: A Long-distance Free-space Implementation Plan of QSDC Without Quantum Memory,” arXiv preprint arXiv:2402.09108, 2024.

\bibitem{b32} A.K. Ekert, “Quantum Cryptography Based on Bell’s Theorem,” \textit{Physical Review Letters}, vol. 67, no. 6, pp. 661-663, 1991.

\bibitem{b33} Y.K. Wong, Y. Zhou, Y.S. Liang, “Quantum Image Denoising with Machine Learning: A Novel Approach to Improve Quan- tum Image Processing Quality and Reliability,” arXiv preprint arXiv:2402.11645, 2024.

\bibitem{b34} Y. Zhou, C.C. Xu, M. Song, Y.K. Wong, K. Du, “A Novel Quantum LSTM Network,” arXiv preprint arXiv:2406.08982, 2024.
\end{thebibliography}
\end{document}